\documentclass[aps,prd,superscriptaddress,showpacs,showkeys,floatfix,preprintnumbers,nofootinbib,letterpaper]{revtex4-2}
\usepackage[Export]{adjustbox}
\usepackage{setspace}
\usepackage{verbatim}
\usepackage{tikz}
\usetikzlibrary {arrows.meta}
\usepackage{color}
\usepackage{amssymb}
\usepackage{pgfplots}
\usepackage{amsmath,amssymb}
\usepackage{mathtools}
\usepackage{hyperref}

\usepackage{graphicx}  
\usepackage{dcolumn}   
\usepackage{bm}        
\usepackage{amssymb}   
\usepackage{physics}
\usepackage{mathtools}
\usepackage{ulem}

\definecolor{Rocco}{rgb}{0.0, 0.5, 0.0}
\definecolor{Resolved?}{rgb}{0.5, 0.0, 0.5}

\newcommand{\DNS}[1]{{\color{orange}#1}}

\newcommand{\QQbar}{{Q\bar Q}}
\newcommand{\Qbar}{{\bar Q}}

\newcommand{\mcD}{{\mathcal D}}
\newcommand{\mcP}{{\mathcal P}}
\newcommand{\mcW}{{\mathcal W}}

\newcommand{\FTEE}{FTE$^2$ }

\newcommand{\la}{\langle}
\newcommand{\ra}{\rangle}

\begin{document}
\title{Entanglement entropy of a color flux tube in (1+1)D Yang-Mills theory}
\author{Rocco Amorosso}
\affiliation{Department of Physics and Astronomy, Stony Brook University, Stony Brook, NY 11794, USA}
\author{Sergey Syritsyn}
\affiliation{Department of Physics and Astronomy, Stony Brook University, Stony Brook, NY 11794, USA}
\author{Raju Venugopalan}
\affiliation{Physics Department, Brookhaven National Laboratory, Upton, NY 11973, USA}
\affiliation{CFNS, Department of Physics and Astronomy, Stony Brook University, Stony Brook, NY 11794, USA}

\begin{abstract}
In recent work~\cite{Amorosso:2024leg}, we computed a novel flux tube entanglement entropy (FTE$^2$) of the color flux tube stretched between a heavy quark-antiquark pair on a Euclidean lattice in (2+1)D Yang-Mills theory. Our numerical results suggested that FTE$^2$ can be partitioned into an internal color entanglement entropy and a vibrational entropy corresponding to the transverse excitations of a QCD string, with the latter described by a thin string model.  
Since the color flux tube does not have transverse excitations in (1+1)D, we analytically compute the contribution of the internal color degrees of freedom to FTE$^2$ in this simpler framework. 
For the multipartite partitioning of the color flux tube, we find the remarkable result that  
FTE$^2$ only depends on the number of times the flux tube crosses the border between two spatial regions, and the dimension of the representation of the color group, but not on the string length.
The result holds independently of whether the branching points are placed on the vertices of the lattice or in the center of plaquettes.
\end{abstract}

\date{\today}

\maketitle
\section{Introduction}
In a recent paper~\cite{Amorosso:2024leg}, we introduced a novel flux tube entanglement entropy (FTE$^2$) defined as 
\begin{equation}
\label{eqn:RenyiDiff}
\tilde{S}^{(q)}_{\vert Q \bar{Q}} = S^{(q)}_{\vert Q \bar{Q}}-S^{(q)} ,
\end{equation}
where $S^{(q)}$ is the order $q$ (R\'{e}nyi) entanglement entropy of gluon fields in
vacuum, and $S^{(q)}_{\vert Q\bar Q}$ is the corresponding quantity in the presence of heavy quark-antiquark sources on a Euclidean lattice in Yang-Mills theory\footnote{One can similarly define FTE$^2$ for the von Neumann entropy by taking the number of replicas $q\rightarrow 1$.}. We showed that FTE$^2$ can be expressed in terms of correlators of Polyakov loops and is therefore manifestly gauge invariant. Further, performing numerical Monte Carlo simulations for SU(2) Yang-Mills theory in (2+1)D, we demonstrated that FTE$^2$ is ultraviolet finite. Our numerical simulations were performed in a ``half-slab" geometry, allowing us to explore the entanglement of a 2-D slab cross-cutting the color flux tube with the region outside it as the width of the slab and its position relative to the heavy-quark sources were varied in the 2-D plane.  

In \cite{Amorosso:2024leg}, we conjectured that \FTEE has two independent contributions,
\begin{equation}
\label{eq:conjecture}
\tilde{S}^{(q)}_{\vert Q \bar{Q}} \stackrel?= {\tilde S}_\text{vibrational} + {\tilde S}_\text{internal}\,,
\end{equation}
where ${\tilde S}_\text{vibrational}\propto\log L/\lambda$ is due to the transverse vibrations of the string of length $L$ with intrinsic wavelength $\lambda$, and 
${\tilde S}_\text{internal}$ is due to internal colorful degrees of freedom of the flux tube. We further postulated that the 
latter should depend only on the number of colors $N_c$ and the number of times $F$ the string crosses the boundary\footnote{The average $\langle F\rangle$ is  taken in the 2+1-D Yang-Mills case because, depending on the 2-D geometry of the $V/\bar V$ partitioning, the color flux tube may or may not intersect the boundaries.} $\partial V$ separating the half-slab region $V$ and its complement $\bar V$,
\begin{equation}
\label{eqn:sun2d_ftee_general_conj}
{\tilde S}_\text{internal} = \langle F \rangle \cdot\log N_c\,.
\end{equation}
We observed in our numerical simulations that the vibrational entanglement entropy is well-described by a thin string model that captures the systematics of these modes for the half-slab geometry as it cross-cuts the flux tube. 
If we assume that Eq.~(\ref{eqn:sun2d_ftee_general_conj}) is robust, our numerical results suggest that \FTEE is 
dominated\footnote{This is the case for short flux tube lengths $L\sqrt{\sigma}\approx 0.6$, where $\sqrt{\sigma}$ is the Yang-Mills string tension.} by ${\tilde S}_\text{internal}$. The validity of Eq.~(\ref{eq:conjecture}) can be tested by performing simulations for $N_c > 2$. 

It is also interesting to ask what happens in (1+1)D where ${\tilde S}_\text{vibrational}=0$ and the only contribution to \FTEE is from Eq.~(\ref{eqn:sun2d_ftee_general_conj}). In this letter, we will compute ${\tilde S}_\text{internal}$ for this case and obtain an analytical result for \FTEE. Specifically, we will show  
that \FTEE for a pair of color sources $Q,\bar Q$ in representation $R_{Q\bar Q}$ is given by 
\begin{equation}
\label{eqn:sun2d_ftee_general}
\tilde{S}^{(q)}_{\vert Q \bar{Q}}\equiv {\tilde S}_\text{internal} = F \cdot \log(\dim R_{Q\bar Q})\,.
\end{equation}
As we will discuss, this analytical (1+1)D result has a number of interesting features that can 
provide useful insight into the higher dimensional computations for $N_c\geq 2$.

While our results for \FTEE in (1+1)D are new, we note that the {\it vacuum} entanglement entropy for (1+1)D Yang-Mills theory on the lattice was previously computed in~\cite{Gromov:2014kia}
and~\cite{Donnelly:2014gva}.  In particular, in the latter work, it was observed that in the non-Abelian case the vacuum entanglement entropy contains an extra term
$\propto F\log\dim R_b$ relative to the Abelian case,
\begin{equation}
\label{eqn:sun2d_vacuum}
S = \sum_{\{R_b\}} p(\{R_b\}) \big[-\log p(\{R_b\}) + F \log\dim \{R_b\} \big]\,,
\end{equation}
where $F$ is the number of boundaries separating $V$ and $\bar V$, similarly to the \FTEE expression, and 
the average is computed over the probability distribution $p(\{R_b\})$ of so-called 
superselection sectors~\cite{Casini:2013rba}.
Each superselection sector corresponds to a unique representation of the color-electric field flux $R_b$ at each point of
the boundary $b\in\partial V$ (also dubbed ``surface charges''), and hence, the contribution of the color degeneracy to the entropy is $\propto \langle\log\dim \{R_b\}\rangle$~\cite{Donnelly:2014gva}.

The form of our \FTEE result in Eq.~(\ref{eqn:sun2d_ftee_general}) indicates that it is produced precisely by this color-entanglement term in Eq.~(\ref{eqn:sun2d_vacuum}) when the gluon fields on the
boundary are modified by adding (a tube of) color-electric flux in representation $R$ connecting the static sources, thereby addressing (in the affirmative) the question of measurability of surface charges raised in \cite{Donnelly:2014gva}.
In particular, the calculation we will present shows that the internal entanglement entropy $S_\text{int}$ is nonzero only for color sources in nontrivial representations ($\dim R>1$) and is thus automatically zero for any Abelian gauge theory. The latter result potentially has important consequences for flux tube phenomenology, especially the validity of the dual-superconductor picture of confinement~\cite{Mandelstam:1974pi,tHooft:1981bkw,Ripka:2003vv}.
The relation between confinement and quantum entanglement in non-Abelian gauge theories has been explored in several papers - see for instance \cite{Klebanov:2007ws,Jokela:2020wgs,Knaute:2024wfh}.

Our analytic calculation is performed using the replica method and yields the same result for both the von Neumann and R\'{e}nyi entanglement entropy for any number of replicas $q$.
The replica method introduces conical singularities at branch cuts which require careful treatment on the lattice.
To ensure that our \FTEE result is not an artifact of our specific prescription for ``cusps,'' conical singularities lying on the intersection
of temporal boundaries between the replicas and the spatial boundary $\partial V$, we consider very distinct discretization setups on Euclidean lattices. In the first of these, the boundaries
between the replicas and regions $V/\bar V$ are described in the ``cusp-on-vertex'' geometry, also studied in \cite{Amorosso:2024leg} (and the majority of prior works referenced therein), where boundaries are drawn through lattice sites and links. We also consider an alternative geometry where the boundaries
between the replicas and regions $V/\bar V$ are drawn through the centers of plaquettes (referred henceforth as ``cusp-in-plaquette''),  In this latter discretization, there are special $q$-sheeted lattice plaquettes that contain the cusps and contain links from all $q$ replicas (total $4q$ links). 
The gauge coupling $\tilde\beta$ on these plaquettes must be different from the coupling $\beta$ on regular plaquettes.
We find that \FTEE computed with the ``cusp-in-plaquette'' discretization is the same  as the ``cusp-on-vertex" \FTEE if the coupling 
$\tilde\beta$ is chosen to be consistent with the weak-coupling (continuum) limit
of $\beta$, which for the latter corresponds to $\beta\to\infty$.

The paper is organized as follows. 
In Section~\ref{sec:app_ent_2d}, we will review the definition of \FTEE on the lattice introduced in \cite{Amorosso:2024leg} and discuss its key features. The computation of  \FTEE on a (1+1)D lattice for the  cusps-on-vertex  case  is discussed in detail in Section~\ref{sec:latt-Vcusp} for any
number of boundaries crossed by the flux tube.
This discussion is generalized in Section~\ref{sec:latt-Pcusp} for the ``cusp-in-plaquette'' discretization, where we discuss further the required modifications
to the gauge action, compute FTE$^2$, and compare our results to those obtained in Sec.\ref{sec:latt-Vcusp}.
For simplicity, our results are obtained for color sources in the fundamental representation $\dim R=N_c$; they   however generalize straightforwardly to any other irreducible representation of the gauge group, with an exception discussed at the end of Section~\ref{sec:latt-Pcusp}.
Our results and their significance are summarized briefly in Section~\ref{sec:conclusions}.

\section{\FTEE in lattice gauge theory
  \label{sec:app_ent_2d}}
In this section, we briefly review the definition of \FTEE in \cite{Amorosso:2024leg}; further motivation, discussions, and results for (2+1)D Yang-Mills theory can be found in this companion paper. 
For Yang-Mills lattice computations, one employs the standard Wilson plaquette gauge action, 
\begin{equation}
\label{eqn:wilson_plaq_action}
S_g[U] = \beta \sum_{x,y, \mu<\nu} \Big[1 - \frac1{N_c}\Re\Tr P_{(x,y),\mu\nu}\Big]\,,
\end{equation}
where $P_{x,\mu\nu} = U_{x,\mu} U_{x+\hat\mu,\nu} U_{x+\hat\nu,\mu}^\dag U_{x,\nu}^\dag$ is the product of links around
the plaquette.
The path integral on an $L_t$-periodic lattice
\begin{equation}
Z = \int \prod_{x,\mu} dU e^{-S_g[U]}\,,
\end{equation}
is performed over all links using the Haar measure $(\int dU = 1)$ and is explicitly gauge-invariant.
Thanks to the gauge symmetry $U_{x,\mu}\to \Omega_x U_{x,\mu}\Omega_{x+\hat\mu}^\dag$ with independent gauge
transformations $\Omega_x$ at any site, this integration can be 
simplified by setting gauge links to unity on any tree of links.
The same holds true for computing integrals with any gauge-invariant link combinations.
The temperature $L_t^{-1}$ can be made arbitrarily small, with the formal expression for the gauge field density matrix given by  
\begin{equation}
\label{eqn:lat_denmat}
\langle U^\text{out} | \rho | U^\text{in}\rangle
  = \frac1{Z} 
    \int _{U(t=0) = U^\text{in}} ^{U(t=L_t) = U^\text{out}} \mcD U e^{-S_g[U]}\,.
\end{equation}
To take the partial trace over gauge fields in region $\bar V$, one integrates over the fields in $\bar V$~\cite{Buividovich:2008kq}:
\begin{equation}
\label{eqn:lat_denmat_reduced}
\langle U_V^\text{out} | \rho_V | U_V^\text{in}\rangle
  = \frac1{Z} 
    \int _{U_V(t=0) = U_V^\text{in}} ^{U_V(t=L_t) = U_V^\text{out}} \mcD U_V  
    \int_{U_{\bar{V}}(t=0) = U_{\bar{V}}(t=L_t)} \mcD U_{\bar{V}}
    \,\, e^{-S_g[U_V, U_{\bar{V}}]}\,,
\end{equation}
This partially-traced density matrix corresponds to the electric-center algebra~\cite{Aoki:2015bsa}.
To calculate the von Neumann entanglement entropy from this quantity, one first calculates the R\'{e}nyi
entropy for an integer number of replicas $q$, and analytically continues to $q\to1$. (This is feasible in 
(1+1)D but not in higher dimensions.)
The entanglement entropy for  $SU(2)$ Yang-Mills theory in (3+1)D was first computed in
\cite{Buividovich:2008kq,Buividovich:2008gq} resulting in the extraction of the entropic $C$-function. 
Several lattice studies have been performed
subsequently~\cite{Velytsky:2008sv,Itou:2015cyu,Rabenstein:2018bri,Rindlisbacher:2022bhe,Bulgarelli:2023ofi,Jokela:2023rba,Ebner:2024mee,Amorosso:2023fzt}.

Our focus here is on the computation of \FTEE on a (1+1)D lattice with infinite spatial size\footnote{We have recovered our analytic results for infinite spatial size on $x$-periodic lattices with sufficiently large  $L_x$.}, as shown
in Fig.~\ref{fig:entent_2d} (right) for $q=2$ and $L_t=3a$ for the case of the flux tube crossing $F=2$ boundaries.
This illustrates the fact that a path traveling around a cusp vertex will cross $4q$ links, unlike a path traveling around a regular vertex.

A color flux tube created by a pair of static quark and antiquark sources is equivalent to adding Polyakov loops to the partition function and the density matrix.
The addition of a color source $Q$ at point $x$ with initial and final color states $i,i'$ results in the modified
density matrix 
\begin{equation}
\label{eqn:lat_denmat_reduced_Q}
\langle U^\text{out},\, x',i' |\rho|U^\text{in},\, x,i\rangle
  \propto \delta_{x', x} \, \big\langle\big[\mcW_{x}\big]_{i' i}\big\rangle_{U^\text{out};U^\text{in}}\,,
\end{equation}
where $\mcW_x = \prod_{\tau=0}^{L_t-1} U_{(x,\tau),\hat t} $ is the product of temporal links
averaged over gauge fields with the same distribution as the pure Yang-Mills density matrix~in Eq.~(\ref{eqn:lat_denmat}).
In \cite{Amorosso:2024leg}, the $Q,\bar Q$ sources were placed in region $\bar V$ at points $x,y$, 
as shown in Fig.~\ref{fig:entent_2d} (right).
In this case, the partial trace of the density matrix can be expressed in terms of the  Polyakov loops  
$\mcP_x = \Tr\mcW_x$ and $\mcP_y^\dag = \Tr\mcW_y^\dag$ as 
\begin{equation}
\label{eqn:lat_denmat_reduced_Qtr_barQtr}
\langle U_V^\text{out}| \rho_{V|Q_x \bar Q_y} |U_V^\text{in}\rangle
  = \frac
   {\big\langle \mcP_x \, \mcP_y^\dag \big\rangle_{U_V^\text{out}; U_V^\text{in}}}
   {\big\langle \mcP_x \, \mcP_y^\dag \big\rangle_{\phantom{U_V^\text{out}; U_V^\text{in}}}}\,,
\end{equation}
where the denominator normalizes the density matrix to unit trace.
If the color sources $Q$ and $\bar Q$ are placed into regions $V$ and $\bar V$, respectively,
the in- and out- states of $\rho_V$ depend on the color indices of the former,
\begin{equation}
\label{eqn:lat_denmat_reduced_Q_barQtr}
\langle U_V^\text{out}, \, x',i'| \rho_{V Q_x|\bar Q_y} |U_V^\text{in}, \, x,i\rangle
  = \frac
   {\big\langle [\mcW_x]_{i' i} \, \mcP_y^\dag \big\rangle_{U_V^\text{out}; U_V^\text{in}}}
   {\big\langle \mcP_y \, \mcP_y\dag \big\rangle_{\phantom{U_V^\text{out}; U_V^\text{in}}}}\,.
\end{equation}

The $q$-th power of the reduced density matrix corresponds to stacking $q$ replicas of the formal 
expressions in Eqs.~(\ref{eqn:lat_denmat_reduced_Qtr_barQtr}) or~(\ref{eqn:lat_denmat_reduced_Q_barQtr}).
When both static quark sources are in $\bar{V}$~\cite{Amorosso:2024leg}, which corresponds to an even number $F$ of boundaries crossed by the QCD string,
\begin{equation}
\Tr \big[ \big( \rho_{V|Q_x \bar Q_y} \big)^q \big]
 = \frac{Z^{(q)}_{|Q\bar Q}}{[Z_{|Q\bar Q}]^q}=\frac {\big\langle \prod_{r=0}^{q-1} \mcP_x^{(r)} \, \mcP_y^{(r)\dag} \big\rangle}
   {\big[\big\langle \mcP_x \, \mcP_y^\dag \big\rangle\big]^q}
    \,\cdot\, \frac{Z^{(q)} }{ Z^q }
   \,,
\end{equation}
where the factor $(Z^{(q)} / Z^q)$ is the ratio of the $q$-replica partition function and the $q$-th power of the regular Yang-Mills vacuum partition function.
It follows from the fact that the $q$-replica Polyakov-loop correlator in the numerator 
$\langle\prod_{r=1}^{q} \mcP_x^{(r)} \, \mcP_y^{(r)\dag}\rangle = Z^{(q)}_{|Q\bar Q} / Z^{(q)}$
is computed on the $q$-replica lattice, 
but the normalization in Eqs.~(\ref{eqn:lat_denmat_reduced_Qtr_barQtr},\ref{eqn:lat_denmat_reduced_Q_barQtr})
(also appearing in the denominator above) is computed on the lattice without replicas, 
$\langle \mcP_x \, \mcP_y^\dag\rangle = Z_{|Q\bar Q} / Z$. Here $Z^{(q)}_{|Q\bar Q}$ is the $q$-replica Yang-Mills partition function in the presence of a  $Q\bar Q$ source, and likewise, $Z_{|Q\bar Q}$ is the corresponding partition function of the full system (the union of $V$ and its complement $\bar V$) on a standard  lattice.

As observed in \cite{Amorosso:2024leg}, UV divergences are now isolated in this factor 
and subtracting it leads to a finite value for FTE$^2$:
\begin{equation}
\begin{aligned}
\label{eqn:RenyiDiffPolyakov_Qtr_barQtr}
\tilde{S}^{(q)}_{\vert Q \bar{Q}} 
  &= S^{(q)}_{\vert Q \bar{Q}} -S^{(q)} 
   = -\frac1{q-1}\log\frac{\Tr\big[\big(\rho_{V|Q\bar Q}\big)^q\big]}{\Tr\big[\big(\rho_V\big)^q\big]}
\\&= -\frac1{q-1}\log\frac{Z^{(q)}_{|Q\bar Q} \, / \, [Z_{|Q\bar Q}]^q} {Z^{(q)} \, / \, [ Z ]^q}
   = -\frac1{q-1}\log\frac{\langle \prod_{r=1}^{q} \mcP_x^{(r)} \, \mcP_y^{(r)\dag}\rangle}
       {\big[\langle \mcP_x \, \mcP_y^\dag\rangle\big]^q}
\,.
\end{aligned}
\end{equation}
 The numerator of the r.h.s in the above equation is the product of Polyakov loops of index $(r)$, over all replicas $1,\cdots, q$.
Note that the ratio of the correlators in Eq.~(\ref{eqn:RenyiDiffPolyakov_Qtr_barQtr}) implicitly subtracts the divergent contribution of the vacuum entanglement entropy $\propto \log[Z^{(q)} / Z^q]$. 

If the $Q$ and $\bar Q$ sources are in the regions $V$ and $\bar V$, respectively, the expression for \FTEE is similar, with the exception that the Polyakov line corresponding to the former must pass through all the replicas:
\begin{equation}
\label{eqn:RenyiDiffPolyakov_Q_barQtr}
\tilde{S}^{(q)}_{Q \vert \bar{Q}} 
  = -\frac1{q-1}\log\frac{\big\langle \Tr\big[\prod_{r=1}^q \mcW_x^{(r)}\big] \,
        \big[\prod_{r=1}^q \mcP_y^{(r)\dag}\big]\big\rangle} {\big[\langle \mcP_x \, \mcP_y^\dag\rangle\big]^q}
\,.
\end{equation}
This correlator needs to be computed in the case of odd number of boundary crossings $F$.

\section{Lattice replicas with cusps on vertices
\label{sec:latt-Vcusp}}

In this section, we will compute \FTEE in the cusp-on-vertex discretization using the lattice definition employed in  \cite{Amorosso:2024leg} and 
outlined in the previous section. We will perform a detailed computation of the \FTEE of region $V$ of a flux tube, with the flux tube crossing the boundary between $V$ and $\bar{V}$ either
$F=2$ or $F=1$ times.
As noted, in the former case, both the $Q$ and $\bar Q$ sources are in $\bar V$, and \FTEE is defined as in 
Eq.~(\ref{eqn:RenyiDiffPolyakov_Qtr_barQtr}). In the latter case, either $Q$ or $\bar Q$ must be in the untraced
region $V$ and Eq.~(\ref{eqn:RenyiDiffPolyakov_Q_barQtr}) defines \FTEE instead.
We will later develop a general scheme for computing multipartite \FTEE for arbitrary $F$ and positions of $Q$ and $\bar Q$.

\subsection{\FTEE with two boundaries}
\begin{figure}[ht!]
\centering
\includegraphics[width=.49\textwidth,valign=c]{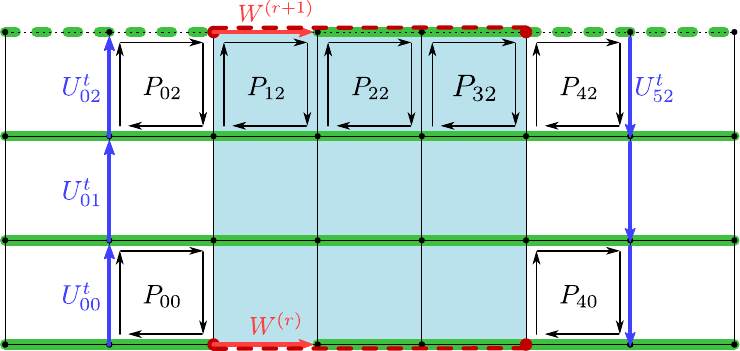}
\includegraphics[width=.4\textwidth,valign=c]{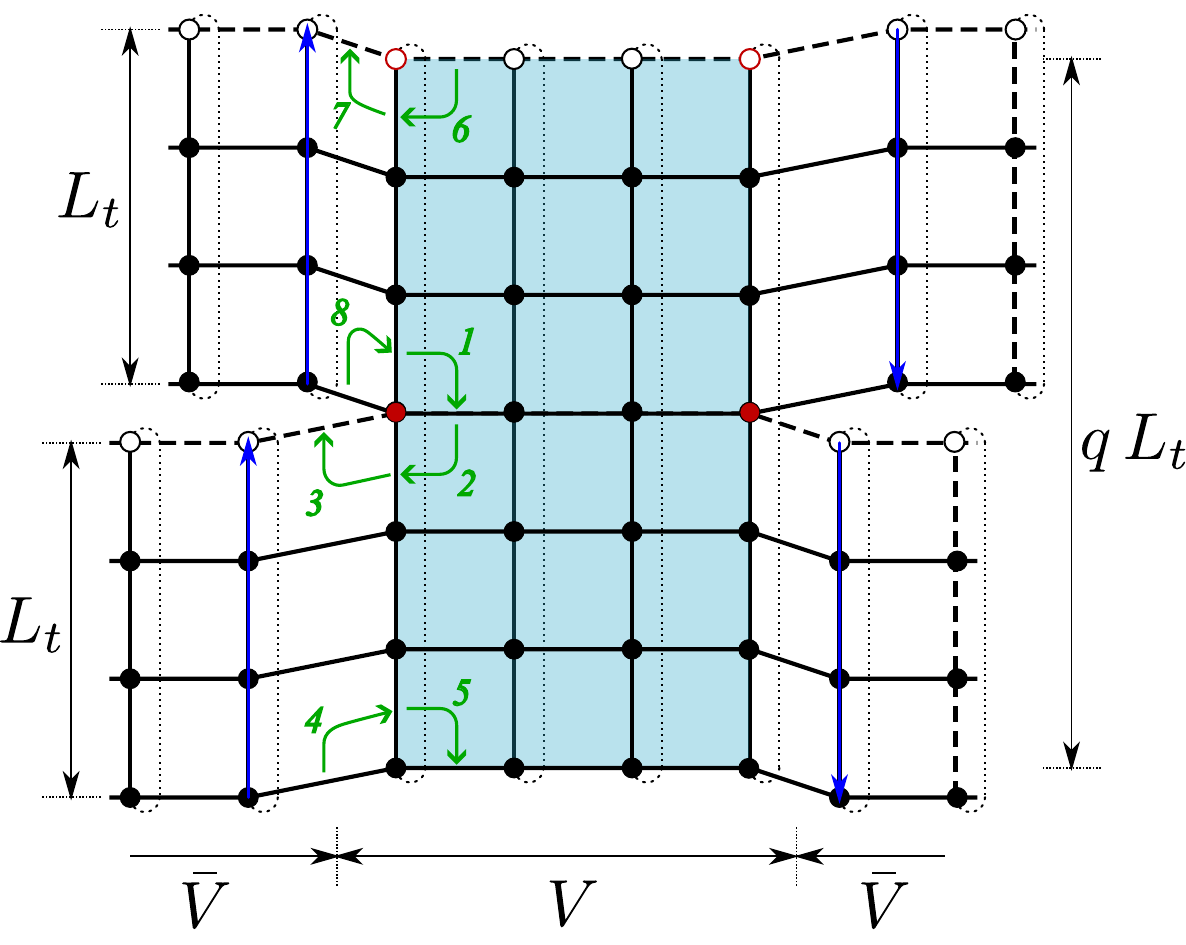}\\
\caption{\label{fig:entent_2d} 
  (Left) Illustration of \FTEE of region $V$ (in cyan) relative to complement regions outside on a 2D lattice. Only one replica $(r=1)$ is shown.
  Spatial links (in bold green) are gauge-fixed to unity.  
  Dashed green lines indicate time-periodic links within the replica (period $L_t=3a$).
  Solid red arrows on the replica boundary between cusps (red filled dots) are not gauge-fixed to unity because of the gauge condition at cusps representing branching points on the multi-sheet $1+1$ manifold. Red dashed
  lines indicate the boundary with the next replica $(r+1)$.
  Blue links denote Polyakov lines for $Q$ and $\bar Q$, and plaquettes $P_{xt}$ are integration variables, equal to link products in the
  indicated order. $\mcW^{(r)}$ and $\mcW^{(r+1)}$ are products of temporal links on adjacent replicas.
  (Right) Geometry of $q=2$ replicas with infinite spatial and $L_t=3a$ temporal extent. 
  Open points on the spacelike edges are periodic wrap-around images of sites connected by dashed  lines.
  Green arrows around one of the cusp sites shows the $q$-sheeted path intersecting $4q$ links.
}
\end{figure}

Consider two static sources, both in $\bar{V}$, separated by spatial distance $L$. A slab corresponding to region $V$ with width $w<L$ sits in between the two sources.  
A single replica of this configuration on a lattice with lattice spacing $a$ is shown in Fig.~\ref{fig:entent_2d} (left), corresponding to $L=5a$, temporal extent $L_t=3a$, and width $w=3a$ of $V$.
The temporal boundaries between replicas and spatial boundaries between regions $V$ and $\bar V$ pass through lattice sites and the links connecting them.
These boundaries intersect at ``cusp'' sites, which will be discussed in more detail below.
In region $V$, the boundary conditions (BC) equate fields on adjacent replicas, so that on the full $q$-replica portions of the lattice
the fields are $(q L_t)$-periodic (not shown), whereas in region $\bar V$ the fields are $L_t$-periodic. 

We will follow here standard approaches to calculate correlators on a lattice~\cite{Gross:1980he,Wadia:1980cp,Creutz:1983njd}.
First, we fix the gauge by setting the spatial links shown (in green) in Fig.~\ref{fig:entent_2d} (left) to unity; 
this has no effect on the final result because the Polyakov-loop correlators in 
Eqs.~(\ref{eqn:RenyiDiffPolyakov_Qtr_barQtr}) and (\ref{eqn:RenyiDiffPolyakov_Q_barQtr}) are gauge-invariant.
For any time slice within a replica, all spatial links can thus be eliminated from the path integral.
However at the temporal boundaries, the gauge transformations $\Omega_{x_C,0}^{(r)}$ are restricted at the ``cusp''
vertices $x_C$ to be periodic~\cite{Aoki:2015bsa},
\begin{equation}
\label{eqn:gauge_cond_replica}
\Omega_{x_C,L_t}^{(r-1)} \equiv \Omega_{x_C,0}^{(r)}
  = \Omega_{x_C,L_t}^{(r)} \equiv \Omega_{x_C,0}^{(r+1)}
\,,\quad r=0\ldots(q-1)
\,,\quad x_C\in\partial V
\,.
\end{equation}
Without this restriction, the plaquettes adjacent to the cusp vertices are not gauge-invariant.
Another way to understand this is to recall that all the $q$ cusp vertices at the same spatial point $x_C$
correspond to the vertex of the conical manifold that defines entanglement as geometric entropy~\cite{Callan:1994py}.
This cone has the central angle $(2\pi q)$ as shown by the path circling cusp vertices in Fig.~\ref{fig:entent_2d} (right).
Because of this, spatial Wilson lines between neighbor cusps cannot be eliminated by gauge fixing, and at least one
gauge link $W^{(r)}$ (shown by the red lines in Fig.~\ref{fig:entent_2d} (left)) must remain unfixed\footnote{
  The gauge can be further fixed by setting to unity $W^{(0)}$, for example, as well as $L_t-1$ temporal links at one spatial point but we skip this to treat all time slices identically.} on each replica boundary $t=rL_t$.
As will be shown below, integrating over these links is necessary to get a nonzero \FTEE value.

In this gauge, all the temporal gauge links $U_{xt}^{\hat t}$ can be expressed through plaquettes $P_{xt}$, one link per time-slice $U_{5t}$, 
and remaining unfixed boundary links $W$ (the Lorentz index $\mu=\hat t$ is omitted below):
\begin{equation}
\begin{array}{lll}
U_{40} = P_{40} U_{50} \,,
& \ldots\,,
& U_{00} = P_{00} P_{10} W^{(r)} P_{20} P_{30} P_{40} U_{50}\,,
\\
U_{41} = P_{41} U_{51}\,,
& \ldots\,,
& U_{01} = P_{01} P_{11} P_{21} P_{31} P_{41} U_{51}\,,
\\
U_{42}  = P_{42} U_{52}\,,
& \ldots \,,
& U_{02} = P_{02} P_{12} P_{22} P_{32} P_{42} U_{52} W^{(r+1)\dag}\,.
\end{array}
\end{equation}
Using the invariance of the Haar measure, integration can be carried out over plaquette variables instead of links.
Such integration is independent for each plaquette $P_{xt}$ as well as the single temporal link on each time-slice $U_{5t}$,
and is performed using the relations~\cite{Creutz:1983njd}
\begin{align}
\label{eqn:group_int_UUdag}
\int dU \, U_{ij} \, (U^\dag)_{kl} &= \frac1{N_c} \delta_{jk} \delta_{il}
\\
\label{eqn:group_int_plaq}
\int dP \, e^{(\beta/N_c) \Re\Tr P} \, P_{ij} &= Z_p(\beta) p(\beta) \delta_{ij}\,,
\end{align}
where the plaquette partition function $Z_p(\beta)$ and its average value $p(\beta)$ are
\begin{equation}
p(\beta) = \frac1{Z_p}\int dP \, e^{(\beta/N_c) \Re\Tr P} \, \big[\frac1{N_c}\Re\Tr P\big]
\,,\quad
Z_p(\beta) = \int dP \, e^{(\beta/N_c) \Re\Tr P}\,.
\end{equation}
The normalization factors $Z_p$ will eventually cancel in all the expressions and thus are ignored below.
Integrating all variables, one time slice at a time, the correlators of the temporal links can be expressed as 
\begin{equation}
\label{eqn:singleT-link-corr}
\begin{aligned}
\la [U_{00}]_{i_0 i_1} [U^\dag_{50}]_{j_1 j_0} \ra_{t=0} 
&= \int dU_{50}\int \prod_x dP_{x0} \, e^{(\beta/N_c)\sum_x\Re\Tr P_{x0}} \, 
     [U_{00}]_{i_0 i_1} [U_{50}^\dag]_{j_1 j_0} 
\\&
  = [p(\beta)]^L \, \frac1{N_c} W^{(r)}_{i_0 j_0 }\delta_{i_1 j_1}\,,
\\
\la [U_{01}]_{i_1 i_2} [U^\dag_{51}]_{j_2 j_1} \ra_{t=1} 
&= [p(\beta)]^L \frac1{N_c}  \delta_{i_1 j_1} \delta_{i_2 j_2}\,,
\\
\la [U_{02}]_{i_2 i_3} [U^\dag_{52}]_{j_3 j_2} \ra_{t=2} 
&
  = [p(\beta)]^L \frac1{N_c} \delta_{i_2 j_2} W^{(r+1)\dag}_{j_3 i_3}\,,
\end{aligned}
\end{equation}
where the time subscripts in $\langle\cdot\rangle_{t}$ indicate that all independent integrals over plaquettes and links on this time-slice have been carried out.
Combining these link averages into the correlator of Polyakov loops within replica $(r)$ by setting $i_0=i_3$ and $j_0=j_3$, we get the partial correlator
\begin{equation}
\label{eqn:singleq-PL-corr}
\la \mcP_0^{(r)} \mcP_L^{(r)\dag} \ra_{t=0\ldots(L_t-1)} = [p(\beta)]^{L_t L} \frac1{N_c} \Tr [W^{(r)} W^{(r+1)\dag}]\,,
\end{equation}
where the subscript $t=0\ldots(L_t-1)$ denotes that the integration takes place only over the $0$th through $(L_t-1)$th time slices. This quantity depends on the boundary links $W$ shared with replicas $(r\pm1)$.
Finally, integrating out the links $W^{(0)}\equiv W^{(q)}, W^{(1)}, \ldots, W^{(q-1)}$ in the product of 
correlators  in Eq.~(\ref{eqn:singleq-PL-corr}) over all replicas yields an additional factor of $1/N_c^{q-2}$,
\begin{equation}
\label{eqn:suN2d_polyakov_loop_corr}
\big\la\prod_{r=0}^{q-1} \mcP_0^{(r)} \mcP_L^{(r)\dag}\big\ra
= \int \prod_{r=0}^{q-1} \Big(d W^{(r)}\, 
    \frac{[p(\beta)]^{L_t L} }{N_c} \Tr\big[ W^{(r)} W^{(r+1)\dag}\big]\Big)
= \frac{[p(\beta)]^{q L_t L} }{N_c^q}\cdot \frac1{N_c^{q-2}}
= \frac{[p(\beta)]^{q L_t L}}{N_c^{2(q-1)}}\,.
\end{equation}
In obtaining this result, we used the fact that in the $(q-1)$ integrals above, 
\begin{equation}  
\int dW^{(r)} \Tr\big[W^{(r-1)}W^{(r)\dag}\big] \Tr\big[W^{(r)}W^{(r+1)\dag}\big] 
=\frac1{N_c} \Tr\big[W^{(r-1)}W^{(r+1)\dag}\big]\,,
\end{equation}
and the result  from the  last integral $\int dW^{(0)} \Tr\big[W^{(0)}W^{(0)\dag}\big]= N_c$, which also follows from Eq. (\ref{eqn:group_int_UUdag}).
For $q=1$, the area law $\la\mcP_0 \mcP_L^\dag\ra = [p(\beta)]^{L_t L}$ 
and the linear $\QQbar$ potential with string tension $a^2\sigma(\beta)=-\log p(\beta)>0$ 
is reproduced.
For $q>1$, the correlator ratio in Eq.~(\ref{eqn:RenyiDiffPolyakov_Qtr_barQtr}) is equal to $1/N_c^{2(q-1)}$ and 
the R\'{e}nyi \FTEE 
\begin{equation}
\label{eqn:sun2d_ftee_G__qbarq}
\tilde S^{(q)}_{|Q\bar Q} = -\frac1{q-1} \,\log\frac
  {\big\la\prod_{r=0}^{q-1} \mcP_0^{(r)} \mcP_L^{(r)\dag}\big\ra}
  {\big(\big\la \mcP_0 \mcP_L^\dag \big\ra\big)^q} 
= \log N_c^2\,.
\end{equation}
Alternatively, analytic continuation to arbitrary $q\to1+\epsilon$ yields the same value for the von Neumann's \FTEE,
\begin{equation}
\tilde S_{|Q\bar Q} = \frac\partial{\partial q} \log \frac 
  {\big\la\prod_{r=0}^{q-1} \mcP_0^{(r)} \mcP_L^{(r)\dag}\big\ra}
  {\big(\big\la \mcP_0 \mcP_L^\dag\big\ra\big)^q} 
= \log N_c^2\,.
\end{equation}

This calculation can be repeated for color sources $Q,\bar Q$ in any irreducible representations $R, \bar R$
of any compact group. The only differences are the dimension of the representation $N_c\to \dim R$ in Eq.~(\ref{eqn:group_int_UUdag}) 
and the plaquette average $p(\beta)\to p_R(\beta)$ for representation $R$ in Eq.~(\ref{eqn:group_int_plaq}).

Notably, this result also indicates that for all Abelian groups, which have only one-dimensional representations, \FTEE is zero. 
This happens because the Abelian links $W^{(r)}$ cancel and do not contribute to the integral,
\begin{equation}
\begin{aligned}
\big\la \prod_{r=0}^{q-1} \la \mcP_0^{(r)} \mcP_L^{(r)\dag} \big\ra_{\rm Abelian}
  = \int\prod_{r=0}^{q-1} dW^{(r)} [ p(\beta) ]^{L_t L} [W^{(r)} W^{(r+1)\dag}]
  = [ p(\beta) ]^{q L_t L} 
  = \big[\big\la \mcP_0 \mcP_L^\dag \big\ra\big]^q\,,
\end{aligned}
\end{equation} 
giving $\tilde S^{(q)}_{|Q\bar Q(\text{Abelian})} = \tilde S_{|Q\bar Q(\text{Abelian})} = 0$.

\subsection{\FTEE with one boundary}
A similar calculation can be carried out with one of the sources (for instance, $Q$) placed in region $V$-the flux tube in this case crosses only one boundary.
The only required modification is in contracting the link correlators  in Eq.~(\ref{eqn:singleT-link-corr}) into the Polyakov loop correlators in Eq.~(\ref{eqn:singleq-PL-corr}).
The Polyakov loop for $\Qbar$ now passes through all the replicas.
The partial average over replica $(r)$ of the untraced Wilson line for $Q$ and the Polyakov loop for $\bar Q$ is
\begin{equation}
\big\la \big[\mcW_0^{(r)} \big]_{ i^{(r)} i^{(r+1)} } \, \mcP_L^{(r)\dag} \big\ra_{t=0\ldots(L_t-1)}
  = [p(\beta)]^{L_t L} \frac1{N_c} \Tr [W^{(r)} W^{(r+1)\dag}]_{i^{(r)} i^{(r+1)}}\,,
\end{equation}
and the average of their product over all replicas is
\begin{equation}
\label{eqn:singleq-PL-corr_2}
\begin{aligned}
\Big\la \Tr \Big[\prod_{r=0}^{q-1} \mcW_0^{(r)} \Big] \, \Big(\prod_{r=0}^{q-1} \mcP_L^{(r)\dag} \Big) \Big\ra 
&= [p(\beta)]^{q L_t L } \frac1{N_c^q} \Tr\Big[\int\prod_{r=0}^{q-1} dW^{(r)} \, 
        W^{(r)} W^{(r+1)\dag} \Big] 
= [p(\beta)]^{q L_t L} \frac1{N_c^{(q-1)}}\,.
\end{aligned}
\end{equation}
Note that in the all-replica average, all the links $W^{(r)}$ cancel, and the overall trace yields only a factor of $\Tr
1 = N_c$.
This is also evident because with only one cusp, all the spatial Wilson lines can be gauge-fixed to unity and
thus integration over the link matrices $W^{(r)}$ cannot therefore contribute to the Polyakov-loop
correlator in  Eq.~(\ref{eqn:singleq-PL-corr_2}). 
Thus the overall factor $\propto 1/N_c^{q-1}$ results in the R\'{e}nyi and von Neumann \FTEE values 
\begin{equation}
\label{eqn:eqn:sun2d_ftee_Gqbar__q}
\tilde S^{(q)}_{Q|\bar Q} = \tilde S_{Q|\bar Q} = \log N_c \,,
\end{equation}
for $Q\in V$ and $\bar Q\in\bar V$ separated by $F=1$ boundary.
The fact that \FTEE in this case is one-half the value in Eq.~(\ref{eqn:sun2d_ftee_G__qbarq}) is due to the flux tube crossing the $V /\bar V$ boundary only once.

\subsection{\FTEE with multiple boundaries}

The previous computations can be repeated for the flux tube crossing multiple boundaries between $V$ and $\bar V$
subregions to show that each boundary contributes a factor of $\log N_c$, 
\begin{equation}
\tilde{S}^{(q)}_{\vert Q \bar{Q}}= 
F\cdot\log N_c\,,
\end{equation}
\begin{figure}[ht!]
\centering
\includegraphics[width=.3\textwidth]{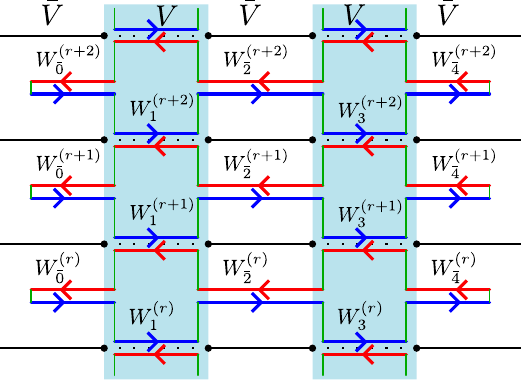}~
\hspace{.15\textwidth}~
\includegraphics[width=.3\textwidth]{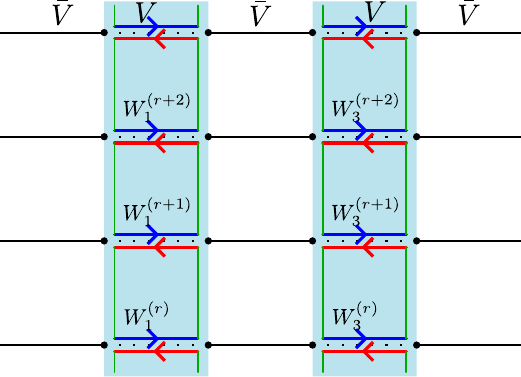}~
\caption{\label{fig:entent_2d_Vcusp_diag_VbarVbar}
  Diagrams representing traces of boundary gauge link matrices $W$ in computing FTE$^2$ between color sources $Q,\bar Q\in\bar V$
  separated by $2z=4$ boundaries. The spatial and temporal/replica directions are horizontal and vertical, respectively.
  (Left) 
  Polyakov-loop correlators in each replica result in
  traces of products of boundary gauge link matrices $W$ (blue arrows) and $W^\dag$ (red arrows).
  Green lines indicate matrix products.
  (Right) 
 Group integration in each replica $r$ results in disjoint traces $\Tr\big[W^{(r)}_\mathcal{B} W^{(r+1)\dag}_\mathcal{B}\big]$ in each of $z$ finite subregions $\mathcal{B}$ of $V$. Further explanation in text. 
}
\end{figure}
\begin{figure}[ht!]
\centering
\includegraphics[width=.4\textwidth]{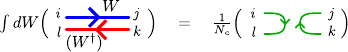}~
\caption{\label{fig:group_int_Vcusp}
  Integration rule for a link matrix bilinear employing Eq.~(\ref{eqn:group_int_UUdag}), which is 
  used to reduce the trace diagrams in Fig.~\ref{fig:entent_2d_Vcusp_diag_VbarVbar}.
}
\end{figure}
as conjectured in Eq.~(\ref{eqn:sun2d_ftee_general}). To prove this result, we need to keep track of the traces that emerge after integrating out
all plaquette and link variables \emph{except} $W^{(r)}$, generalizing Eq.~(\ref{eqn:singleq-PL-corr}).
These spatial links appear between any pair of adjacent cusps, and as before, the final integration over $W^{(r)}$ will
``stitch together'' the replicas.
To illustrate this, and to organize counting of the $N_c$ factors, we represent the traces of the remaining gauge
matrices $W^{(r)}$ diagrammatically in Fig.~\ref{fig:entent_2d_Vcusp_diag_VbarVbar},
where we first consider the case $Q,\bar Q\in \bar V$ with the flux tube crossing $2z$ boundaries (for $z=2$) while traversing
$z$ disjoint subregions of $V$ and $z-1$ subregions of $\bar V$ in between.
Only the remaining unintegrated links $W^{(r)}$ and  $W^{(r)\dag}$ are shown by the blue and red arrows, respectively.
In strips of region $\bar V$, the links are periodic within a replica, indicated by the solid black lines
representing partial trace $\rho_V=\Tr_{\bar V}\rho$. The $W^{(r)}_{\bar0,\bar2,\ldots}$ links are drawn inside the temporal extent of each replica and strip to indicate that they are unique to them.
Conversely, $W^{(r)}_{1,3,\ldots}$ in region $V$ are shared by adjacent replicas and are shown close to their
temporal boundaries.
The dotted lines indicate dot-products of partially-traced density matrices in $\Tr[(\rho_V)^q]$.
For generality we also include the links $W^{(r)}_{\bar0}$ (between left infinity and the leftmost cusp) as
well as $W^{(r)}_{\overline{2z}}$ (from rightmost cusp to right infinity).

Since each Polyakov line is traced within a replica, their correlators result in traces of
counter-clockwise products similar to Eq.~(\ref{eqn:singleq-PL-corr}),
\begin{equation}
\label{eqn:singleq-PL-corr_multi}
\la \mcP_0^{(r)} \mcP_L^{(r)\dag} \ra_{t=0\ldots(L_t-1)} 
= \frac{[p(\beta)]^{L_t L}}{N_c} \int
  \underbrace{dW_{\bar0}\dotsm dW_{\overline{2z}}}_{(z+1)\text{ integrals}}\,
  \frac1{N_c} \Tr\big[
  \underbrace{W^{(r)}_{\bar0} W^{(r)}_{1} \ldots  W^{(r)}_{2z-1} W^{(r)}_{\overline{2z}}}_{
      (2z+1)\text{ matrices}} \cdot 
  \underbrace{W^{(r)\dag}_{\overline{2z}} W^{(r+1)}_{2z-1} \ldots W^{(r+1)\dag}_{1} W^{(r)\dag}_{\bar0}}_{
      (2z+1)\text{ matrices}}
  \big]\,,
\end{equation}
where labels $\bar{\mathcal{A}}=\bar0,\bar2\ldots\overline{2z}$ and $\mathcal{B}=1,3\ldots2z-1$ enumerate alternating subregions (strips) of $\bar V$ and $V$ respectively.
Note that the left- and right-most gauge link matrices $W^{(r)}_{\bar1}$ and $W^{(r)}_{\overline{2z}}$ cancel and do not contribute to the final result. (Alternatively, they can be gauge-fixed to unity because they are not between cusps.) Integrating over the other $(z-1)$ gauge link matrices $W^{(r)}_{\bar2},\ldots,W^{(r)}_{\overline{2(z-1)}}$ in subregions
$\bar V$ with the help of Eq.~\ref{eqn:group_int_UUdag}, as visualized in Fig.~\ref{fig:group_int_Vcusp}, 
we get a factor of $(1/N_c)^{z-1}$ and separate traces in each of $z$ subregions $V$-- as shown in 
Fig.~\ref{fig:entent_2d_Vcusp_diag_VbarVbar} (right): 
\begin{equation}
\la \mcP_0^{(r)} \mcP_L^{(r)\dag} \ra_{t=0\ldots(L_t-1)} 
  = [p(\beta)]^{L_t L} \frac1{N_c^z} \underbrace{
  \Tr\big[W^{(r)}_{1} W^{(r+1)\dag}_{1}\big] \dotsm
  \Tr\big[W^{(r)}_{2z-1} W^{(r+1)\dag}_{2z-1}\big]}_{(z)\text{ traces}} \,.
\end{equation}
The integration over the remaining $(z\times q)$ gauge link matrices $W^{(r)}_1,\ldots,W^{(r)}_{2z-1}$ yields another
factor $(1/N_c)^{z(q-2)}$ similarly to Eq.~(\ref{eqn:suN2d_polyakov_loop_corr}), giving 
\begin{equation}
\big\la\prod_{r=0}^{q-1} \mcP_0^{(r)} \mcP_L^{(r)\dag}\big\ra
  = \int \prod_r dW^{(r)}_1 \ldots dW^{(r)}_{2z-1}  
    \big\la \mcP_0^{(r)} \mcP_L^{(r)\dag} \big\ra_{t=0\ldots(L_t-1)}
  = \frac{[p(\beta)]^{q L_t L}}{N_c^{zq}}\cdot\frac1{N_c^{z(q-2)}}
  = \frac{[p(\beta)]^{q L_t L}}{N_c^{2z(q-1)}}\,.
\end{equation}
so that \FTEE is indeed $\tilde{S}^{(q)}_{\vert Q \bar{Q}} = (2z)\log N_c$.
\begin{figure}[ht!]
\centering
\includegraphics[width=.24\textwidth]{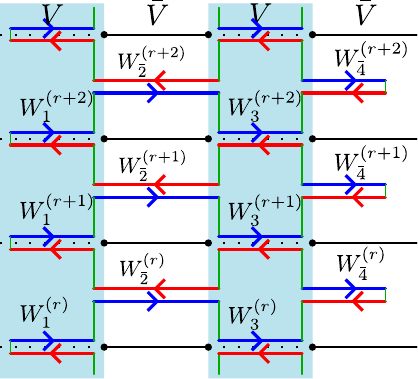}~
\hspace{.15\textwidth}~
\includegraphics[width=.24\textwidth]{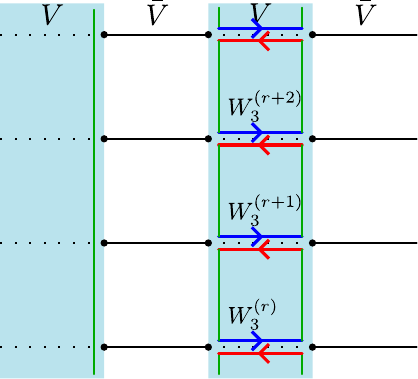}~
\caption{\label{fig:entent_2d_Vcusp_diag_VbarV}
   $W$-trace integration 
   with color sources
  $Q\in V$, $\bar Q\in \bar V$, with flux tube crossing $(2z-1)=3$ boundaries.
  (Left) Polyakov loop for $Q$ being in $V$ 
  results in a single trace through all  replicas $r=0\ldots(q-1)$. 
  (Right) Integrating out gauge link matrices $W^{(r)}_{\bar{\mathcal{A}}}$ in all subregions $\bar V$ of each replica $(r)$
  results in a disjoint trace $\Tr\big[W^{(r)}_\mathcal{B} W^{(r+1)\dag}_\mathcal{B}\big]$ in each of $(z-1)$ finite subregions $\mathcal{B}$
  of $V$ and additional factor $\Tr1=N_c$.
}
\end{figure}
The computation with both sources $Q$ and $\bar Q$ in $V$ instead of $\bar V$ leads to a trace diagram analogous to
Fig.~\ref{fig:entent_2d_Vcusp_diag_VbarVbar} and is completely identical,  if one switches 
$V$ and $\bar V$.

However, as we discussed previously for the $z=1$ case, if $Q$ is in region $V$ and $\bar Q$ is in region $\bar V$, the topology of the trace diagram is different.  For $z=2$, this is shown in Fig.~\ref{fig:entent_2d_Vcusp_diag_VbarV}. As indicated, the Polyakov loop for $\bar Q$ results in a single trace ``snaking'' through all the $r=0\ldots(q-1)$ replicas.
Nevertheless, the integration over link matrices $W^{(r)}$ is very similar.
Firstly, the integrals over each of the $(z-1)\times q$ matrices $W^{(r)}_{\bar2},\ldots,W^{(r)}_{\overline{2z-2}}$ 
in $\bar V$ yields a factor of $(1/N_c)^{(z-1)q}$.
This leaves $(z-1)\times q$ separate traces of link matrix products $W^{(r)}_3,\ldots,W^{(r)}_{2z-1}$ in $V$,
as well as a factor of $\Tr 1 = N_c$, analogously to the discussion leading from
Eq.~(\ref{eqn:singleq-PL-corr_2}) to Eq.~(\ref{eqn:eqn:sun2d_ftee_Gqbar__q}).
Finally, integrating over the remaining $(z-1)$ matrices $W^{(r)}_{\mathcal{B}}$ in each replica yields a factor of 
$(1/N_c)^{(z-1)(q-2)}$, giving 
\begin{equation}
\begin{aligned}
\la \mcP_0 \mcP_L^\dag \ra
&= \frac{[p(\beta)]^{q L_t L}}{N_c^q} \cdot \frac1{N_c^{(z-1)q}}\cdot \Tr1 \, \cdot \, 
\prod_r \underbrace{
\Big(\int dW^{(r)}_{\mathcal{B}} \, 
  \Tr\big[W^{(r)}_{\mathcal{B}} W^{(r+1)\dag}_{\mathcal{B}}\big]\Big)\,\dotsm}_{(z-1)\text{ factors}}
\\&= \frac{[p(\beta)]^{q L_t L}}{N_c^q} \cdot \frac1{N_c^{(z-1)q}} \cdot N_c \cdot
    \frac1{N_c^{(z-1)(q-2)}}
  = \frac{[p(\beta)]^{q L_t L}}{N_c^{(2z-1)(q-1)}}\,,
\end{aligned}
\end{equation}
demonstrating that \FTEE $\tilde{S}^{(q)}_{\vert Q \bar{Q}} = (2z - 1)\log N_c$ is proportional to $(2z-1)$, or equivalently, in this general case as well, it is proportional to the number times the flux tube crosses
the $\bar V / V$ boundary.  This completes the proof of Eq.~(\ref{eqn:sun2d_ftee_general}) for the ``cusp-on-vertex" geometry.

\section{Lattice replicas with cusps in plaquette centers
\label{sec:latt-Pcusp}}


\begin{figure}[ht!]
 \centering
\includegraphics[width=.49\textwidth,valign=c]{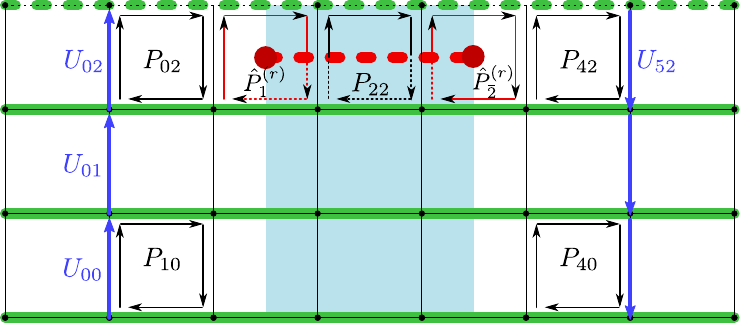}
\includegraphics[width=.4\textwidth,valign=c]{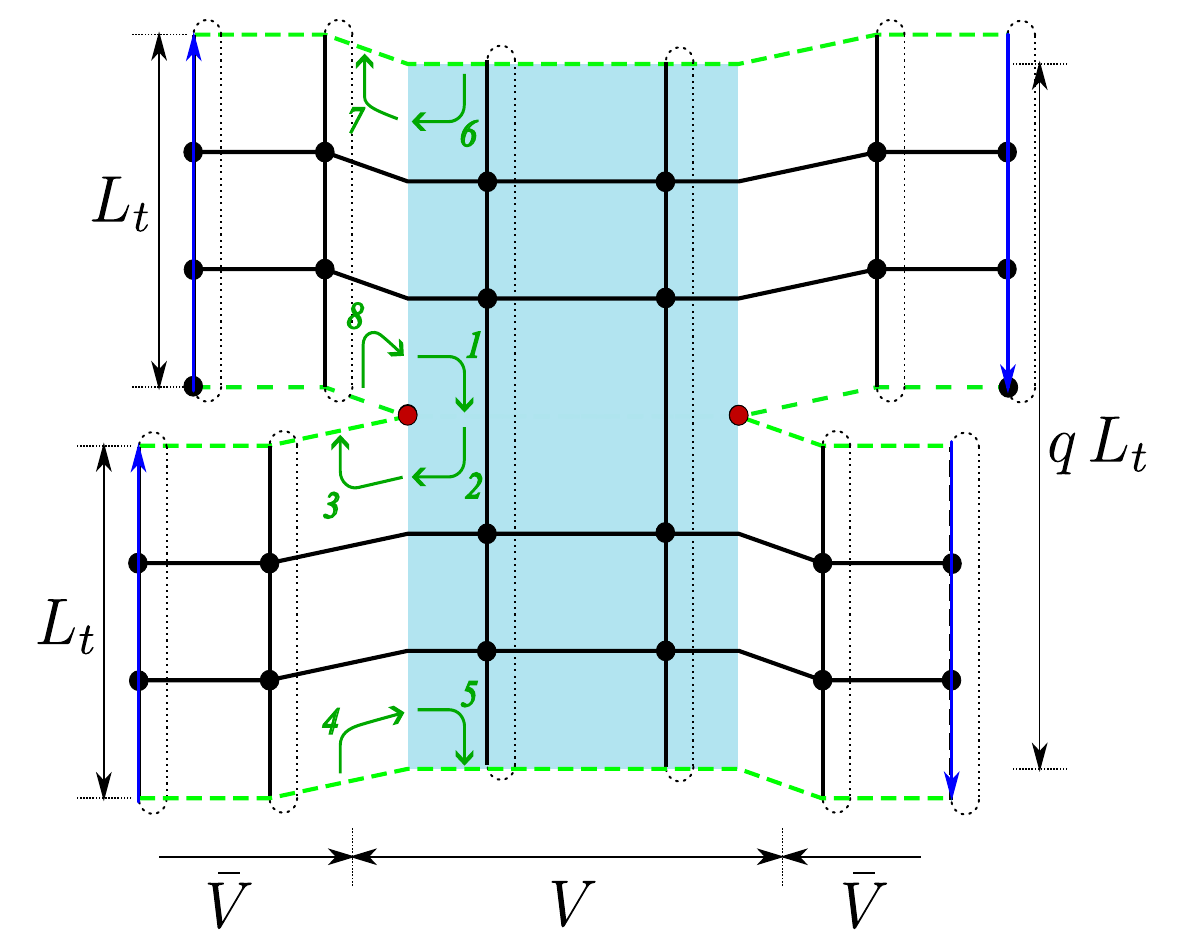}\\
\caption{\label{fig:entent_2d_Pcusp}
  (Left)  \FTEE with cusps (red dots) in center of boundary plaquettes. Spatial links (in green) are gauge-fixed to unity. 
  Black and green dashed lines indicate periodic boundary conditions in time within one replica. Red boundary plaquettes belong to 
  ``big plaquettes'', 
  with dashed links belonging
  to replica $(r-1)$.
 Red dashed line denotes replica boundary.
  (Right) Geometry of $q=2$ replicas with infinite spatial and $L_t=2$ temporal extent. 
  Green arrows around one of the cusps shows $q$-sheeted path intersecting $4q$ links. 
  Green dashed lines bisect plaquettes, denoting periodic boundary conditions on the lattice.
}
\end{figure}

Another choice for discretizing the multi-sheet replica manifold that defines \FTEE
is to draw the temporal boundaries between replicas and spatial boundaries between regions
$V,\bar V$ through plaquette centers as shown in Fig.~\ref{fig:entent_2d_Pcusp}.
In this case, the ``cusps'' will be located in the centers of multi-sheet ``big plaquettes''~\cite{Chen:2015kfa}.
The gauge condition (Eq.~(\ref{eqn:gauge_cond_replica})) is no longer present because there are no Wilson lines that can end 
on cusps.
As before, we consider first the case $Q,\bar Q\in \bar V$ separated by two boundaries with one segment of $V$.
The gauge action on an entire $q$-replica lattice has now contributions from two types of plaquettes,
\begin{equation}
\label{eqn:action_Pcusp}
S' = \beta\sum_{P}{}^\prime \big[1-\frac1{N_c}\Re\Tr P\big] 
  + \tilde\beta_q\sum_{\mathbb{P}}\big[1 - \frac1{N_c}\Re\Tr\mathbb P\big]\,,
\end{equation}
where the first sum is over all plaquettes (on all replicas) with the exception of plaquettes containing the cusp, 
while the second sum is over $q$-sheet ``big plaquettes'', one per each $V/\bar V$ boundary.
A ``big plaquette'' is a Wilson loop made of $(2q)$ temporal and $(2q)$ spatial links winding $q$ times around a cusp and shared by all replicas:
\begin{align}
\label{eqn:bigplaq_Pcusp}
\mathbb{P}_1
  = \underbrace{U_{12}^{(q-1)} U_{22}^{(q-1)\dag}}_{\hat P_1^{(q-1)}} 
    \underbrace{U_{12}^{(q-2)} U_{22}^{(q-2)\dag}}_{\hat P_1^{(q-2)}} \ldots 
    \underbrace{U_{12}^{(0)}   U_{22}^{(0)\dag}  }_{\hat P_1^{(0)}} \,,\\
\mathbb{P}_{\bar2}^\dag
  = \underbrace{U_{42}^{(q-1)} U_{32}^{(q-1)\dag}}_{\hat P_{\bar2}^{(q-1)\dag}} 
    \underbrace{U_{42}^{(q-2)} U_{32}^{(q-2)\dag}}_{\hat P_{\bar2}^{(q-2)\dag}} \ldots 
    \underbrace{U_{42}^{(0)  } U_{32}^{(0)\dag}  }_{\hat P_{\bar2}^{(0)\dag}} \,.
\end{align}
The $\hat P_{1,\bar2}^{(r)}$ variables above are partial plaquettes defined as clockwise products of links
starting at the lower left corner.
Note that  $\hat P_{1}^{(r)}$ starts in replica $r$ and ends in $(r-1)$, 
while $\hat P_{\bar2}^{(r)}$ starts in $r$ and ends in $(r+1)$.
The same would be true for any Wilson line crossing the boundary, 
for example going up from replica $(r)$ leads into replica $(r+1)$.
The links in $V$ on the time slices containing cusps belong to both adjacent replicas, 
and are labeled above by the replica containing their end-site.
The big-plaquette coupling $\tilde\beta_q$ describes the interaction near cone singularities (cusps) and needs to be
selected using some reasonable prescription leading to the correct continuum limit as $\beta\to\infty$.
For example, matching in the weak coupling limit by setting $\tilde\beta_q=\beta/q$, also employed in \cite{Chen:2015kfa}, 
leads to \FTEE consistent with Sec.~\ref{sec:latt-Vcusp}.

The transfer matrix construction of entanglement entropy in \cite{Aoki:2015bsa} can be repeated 
for the cusp-on-plaquette (1+1)D case by using timelike links (instead of spacelike links) as the degrees of freedom. 
In the axial gauge with spatial links gauge-fixed to the identity ($A_1=0$), plaquettes depend only on timelike links $U_{x,t}$.
The transfer matrix can then be written as
\begin{equation}
    \langle U | \hat{T} | U' \rangle=\text{exp}[-\frac{1}{2}S_0(U)]\text{exp}[-\frac{1}{2}S_0(U')]\,\qquad{\rm with}\qquad
    S_0(U)=-\frac{\beta}{N_c}\sum\limits_{x} \text{Re}[\text{Tr}[U_{x} U^{\dagger}_{x+1}]]\,.
\end{equation}
However the center algebra in such a construction is not clear, and especially so for $D>2$.
It is nevertheless instructive to examine \FTEE in this discretization and verify that it leads to the same result.

Unlike \FTEE for the cusp-on-vertex geometry where spatial
gauge links $W^{(r)}$ on the replica boundaries cannot be gauge-fixed to unity (and have to be integrated separately), cusp-in-plaquette correlators of timelike links on time-slices with cusps depend on partial plaquettes $\hat P$ that contribute to the action only within the big plaquettes $\mathbb{P}_{1,\bar2}$
\begin{equation}
\begin{aligned}
\label{eqn:singleT-link-corr-Pcusp}
\la [U_{02}]_{i_2 i_0} [U^\dag_{52}]_{j_0 j_2} \ra_{t=2} 
&\propto [p(\beta)]^{L-2} \int dU_{52} \, 
[\hat P_1^{(r)} \hat P_{\bar2}^{(r)} U_{52} ]_{i_2 i_0} [U_{52}]^\dag_{j_0 j_2} 
\\&\propto [p(\beta)]^{L-2} \frac1{N_c} [\hat P_1^{(r)} \hat P_{\bar2}^{(r)}]_{i_2 j_2} \delta_{i_0 j_0} \,,
\end{aligned}
\end{equation}
where the periodic boundary conditions $i_3\equiv i_0$, $j_3\equiv j_0$ are taken into account,
and the integral over the big-plaquette factors $P_{1,\bar2}^{(r)}$ taken last. 
Assembling the link correlators in Eqs.~(\ref{eqn:singleT-link-corr}) and~(\ref{eqn:singleT-link-corr-Pcusp}) into the Polyakov-loop correlator,
\begin{equation}
\label{eqn:singleq-PL-corr_Pcusp}
\la \mcP_0^{(r)} \mcP_L^{(r)\dag}\ra_{t=0\ldots(L_t-1)} 
  = \frac{[p(\beta)]^{L_t L - 2}}{N_c} \Tr [\hat P_1^{(r)} \hat P_{\bar2}^{(r)}]
\end{equation}
and integrating out the cusp plaquettes $\hat P^{(r)}_{1,\bar2}$ 
yields the $q$-replica average
\begin{equation}
\label{eqn:QQbar_Pcusp_VbarVbar}
\begin{aligned}
\la \prod_{r=0}^{q-1} \mcP_0^{(r)} \mcP_L^{(r)\dag}\ra
  &=\frac{p^{q(L_t L -2)}}{N_c^q} 
  \int \Big\{\prod_{r=0}^{q-1} d\hat P_1^{(r)} d\hat P_{\bar2}^{(r)}\,\Tr[\hat P_1^{(r)} \hat P_{\bar2}^{(r)}] \Big\} \,
  e^{(\tilde\beta_q / N_c)\Re\Tr(\mathbb{P}_1+\mathbb{P}_{\bar2})}
\\&=\frac{p^{q(L_t L -2)}\tilde p_q^2}{N_c^q} 
  \int \Big\{\prod_{r=0}^{q-2} d\hat P_1^{(r)} d\hat P_{\bar2}^{(r)}\,\Tr[\hat P_1^{(r)} \hat P_{\bar2}^{(r)}] \Big\} \,
  \Tr[ \hat P_1^{(0)\dag} \ldots \hat P_1^{(q-2)\dag} \hat P_{\bar2}^{(q-2)\dag} \ldots \hat P_{\bar2}^{(0)\dag} ]
\\&=\frac{p^{q L_t L}\cdot(\tilde p_q/p^q)^2}{N_c^q}
     \cdot\frac1{N_c^{q-1}}\cdot\Tr 1
  =\frac{p^{q L_t L}\cdot(\tilde p_q/p^q)^2}{N_c^{2(q-1)}} 
\,.
\end{aligned}
\end{equation}
The second line is obtained by integrating over $d\mathbb{P}_{1,\bar2}$ after substitution
$\hat P_1^{(q-1)} = \mathbb{P}_1 \hat P_1^{(0)\dag}\dotsm\hat P_1^{(q-2)\dag}$,
$\hat P_{\bar2}^{(q-1)} = \hat P_{\bar2}^{(q-2)\dag}\dotsm\hat P_{\bar2}^{(0)\dag}\mathbb{P}_{\bar2}$,
which also yields a factor $\tilde p_q = p(\tilde\beta_q)$ for each $V/\bar V$ boundary.
Finally, integrals over $P^{(r)}_{1,\bar2}$, $r=(q-2),\ldots,0$ are carried out using 
Eq.~(\ref{eqn:group_int_UUdag}) and yield overall factor $(1/N_c)^{q-1}$ as well as $\Tr 1=N_c$.
Note that the usual Polyakov-line correlator ($q=1$) does not have factors of $\tilde p_q$, so the entanglement
entropy will contain a term $\propto\log (\tilde p_q / p^q)$ 
\begin{equation}
\tilde S^{(q)}_{|Q\bar Q} = -\frac1{q-1}
    \log\frac{\la \prod_{r=0}^{q-1} L_Q^{(r)} L_{\Qbar}^{(r)}\ra}
             {[\la L_Q L_{\Qbar}\ra]^q} 
= \log N_c^2 -\frac2{q-1} \log\frac{\tilde p_q}{p^q}\,.
\end{equation}
In the continuum limit where $\beta\to\infty$, $p\to1$, and $a^2\sigma=-\log p\to0$, it is reasonable to expect that also
$\tilde\beta_q\to\infty$ and $\tilde p_q\to 1$, so that the extra term $\propto\log(\tilde p_q / p)$ vanishes.
At a finite $\beta\gg1$ (weak coupling), this extra term is small if the ``big-plaquette'' coupling is set to $\tilde\beta_q=\beta/q$~\cite{Chen:2015kfa} since
\begin{equation}
\label{eqn:tilbeta_tuning}
p(\beta/q) = [p(\beta)]^q + O(\beta^{-2}) \,,
\end{equation}
and the R\'{e}nyi and the von Neumann \FTEE are equal to $\log N_c^2$ up to a correction $\propto(1/\beta^2)$.

\begin{figure}[ht!]
\centering
\includegraphics[width=.3\textwidth]{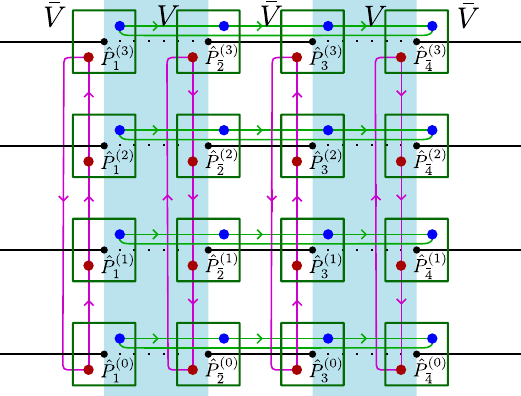}~
\hspace{.15\textwidth}~
\includegraphics[width=.30\textwidth]{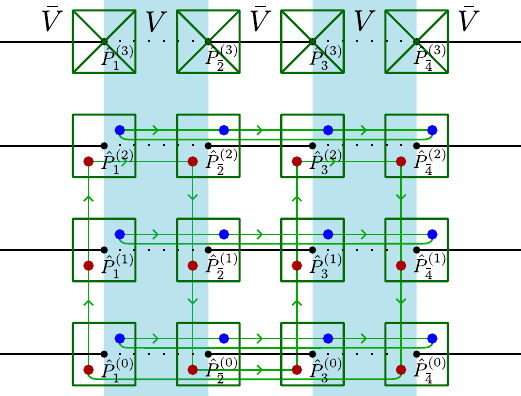}~
\caption{\label{fig:entent_2d_Pcusp_diag_VbarVbar}
  Diagrams of traces of partial plaquette variables $P^{(r)}_{a,...}$  in computing \FTEE between color sources $Q,\bar Q\in\bar V$ 
  crossing $2z=4$ boundaries.
  Replica boundaries are similar to Fig.~\ref{fig:entent_2d_Vcusp_diag_VbarVbar}.
  Plaquette variables $P$ are represented by a blue dot ($P^\dag$ by a red dot).
  Green lines represent traces of their products in the order around the loop, and the magenta lines are the
  big-plaquette 
  contributions to the action. 
  (Left) Integrating out link $U$ and non-cusp plaquette $P$ matrices, 
  Polyakov-loop correlators 
  results in a separate ``horizontal'' trace in each replica.
  Green lines indicate matrix products and magenta lines indicate the traces of large 
  plaquettes.
  (Right) Trace diagram after integrating top row of plaquette matrices.
}
\end{figure}
Similarly to the previous section, the calculation in Eq.~(\ref{eqn:QQbar_Pcusp_VbarVbar}) can be generalized 
for a flux tube crossing multiple alternating subregions $V$ and $\bar V$.
First, we study the case when $Q,\bar Q$ are in $\bar V$ and the flux tube crosses $2z$ boundaries.
To visualize this calculation, we use trace diagrams in Fig.~\ref{fig:entent_2d_Pcusp_diag_VbarVbar}~(left) generalizing 
the first line of Eq.~(\ref{eqn:QQbar_Pcusp_VbarVbar}). 
Polyakov loops on each replica produce $q$ ``horizontal'' traces 
$\Tr\big[ \hat P^{(r)}_1 \hat P^{(r)}_{\bar2}\dotsm\big]$ 
(green loops), while the action depends on $(2z)$ ``vertical'' traces 
$\Re\Tr\big[\hat P^{(0)\dag}_1 \dotsm \hat P^{(q-1)\dag}_1\big]$,
$\Re\Tr\big[\hat P^{(q-1)\dag}_{\bar2} \dotsm \hat P^{(0)\dag}_{\bar2}\big]$, $\ldots$
(magenta loops).
Note that the order of the plaquette matrices in the ``vertical'' traces depends on which boundary the flux tube 
crosses, $\bar V$ to $V$ or vice versa, as it goes $Q\to\bar Q$, in accordance with Eq.~(\ref{eqn:bigplaq_Pcusp}).
Integrating out the top row of plaquette matrices $\hat P^{(q-1)}_1, \hat P^{(q-1)}_{\bar2},\ldots$ yields $(2z)$ 
factors of $\tilde p_q$ and the diagram in Fig.~\ref{fig:entent_2d_Pcusp_diag_VbarVbar}~(right).
The top-row ``horizontal'' trace is replaced with the trace of the ``meandering'' product of matrices
$P^{(r)\dag}_{1,\bar2,\ldots}$, $r=0,\ldots(q-2)$, 
similarly to the second line of Eq.~(\ref{eqn:QQbar_Pcusp_VbarVbar}).
\begin{figure}[ht!]
\centering
\includegraphics[width=.4\textwidth]{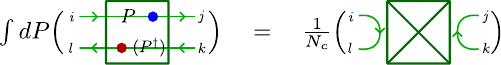}~
\caption{\label{fig:group_int_Pcusp}
  Integration rule employing the link matrix bilinear in Eq.~(\ref{eqn:group_int_UUdag}) 
  used to reduce the trace diagrams shown in Fig.~\ref{fig:entent_2d_Pcusp_diag_VbarVbar}).
}
\end{figure}
\begin{figure}[ht!]
\centering
\includegraphics[width=.30\textwidth]{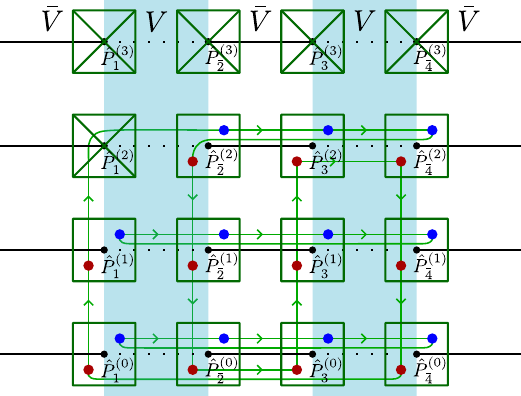}~
\hspace{.15\textwidth}~
\includegraphics[width=.30\textwidth]{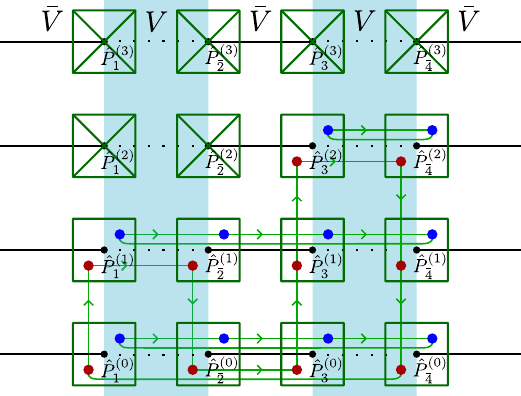}~
\caption{\label{fig:entent_2d_Pcusp_diag_VbarVbar2}
  Two steps of diagram reduction equivalent to integrating out plaquette matrices $P^{(2)}_1$ and
  $P^{(2)}_{\bar2}$.
}
\end{figure}

Further integration over the plaquettes is carried out using Eq.~\ref{eqn:group_int_UUdag}.
The corresponding diagram reduction rule is shown in Fig.~\ref{fig:group_int_Pcusp}.
Two initial steps of reducing the trace diagram are shown in Fig.~\ref{fig:entent_2d_Pcusp_diag_VbarVbar2}.
Integrating the entire row of plaquettes yields the factor $(1/N_c)^{2z-1}$.
Carrying this out for all $(q-1)$ rows, and keeping in mind the last trace factor $\Tr 1=N_c$, we get
the all-replica Polyakov loop correlator
\begin{equation}
\big\la\prod_r \mcP_0^{(r)} \mcP_L^{(r)\dag}\big\ra 
  = \frac{p^{q L_t L}\cdot(\tilde p_q/p^q)^{2z}}{N_c^q} \cdot\frac1{N_c^{(2z-1)(q-1)}}\cdot N_c
  = \frac{p^{q L_t L}\cdot(\tilde p_q/p^q)^{2z}}{N_c^{2z(q-1)}}
\end{equation}
that gives the \FTEE $\tilde S^{(q)}_{|Q\bar Q}=(2z)\log N_c$ consistent with calculation in the cusp-on-vertex replica geometry upon the
condition $\tilde p_q\to p^q$ as before.

\begin{figure}[ht!]
\centering
\includegraphics[width=.24\textwidth]{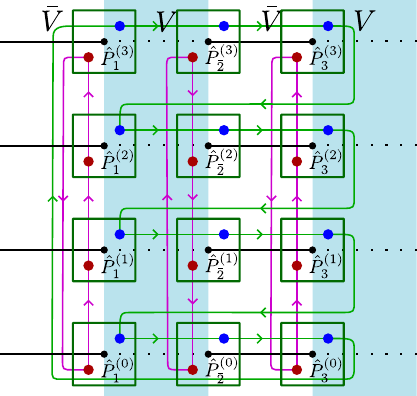}~
\hspace{.15\textwidth}~
\includegraphics[width=.24\textwidth]{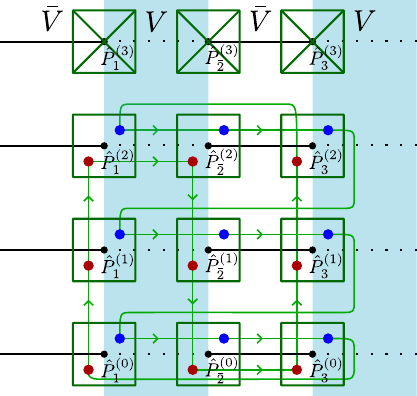}~
\caption{\label{fig:entent_2d_Pcusp_diag_VbarV}
  Diagrams for computing \FTEE with $Q\in\bar V$ and $\bar Q\in V$ with the flux tube crossing $(2z-1)$ boundaries.
}
\end{figure}
The calculation of \FTEE in the case $Q\in\bar V$ and $\bar Q\in V$ with the flux tube crossing $(2z-1)$ boundaries is similar, with the diagrams shown in Fig.~\ref{fig:entent_2d_Pcusp_diag_VbarV}.
Note that the link correlator~(\ref{eqn:singleT-link-corr-Pcusp}) must be replaced with
\begin{equation}
\label{eqn:singleT-link-corr-Pcusp_VbarV}
\big\la [U_{02}]_{i^{(r)}_2 i^{(r)}_0} [U^\dag_{52}]_{j^{(r)}_0 j^{(r-1)}_2} \big\ra_{t=2} 
\propto [p(\beta)]^{L-2} \frac1{N_c} 
  [\hat P_1^{(r)} \hat P_{\bar2}^{(r)}\dotsm]_{i^{(r)}_2 j^{(r-1)}_2} 
  \delta_{i^{(r)}_0 j^{(r)}_0} \,,
\end{equation}
which results in the single trace with the diagram shown in Fig.~\ref{fig:entent_2d_Pcusp_diag_VbarV} (left).
Integrating out the top row of plaquette matrices yields the complicated trace diagram in
Fig.~\ref{fig:entent_2d_Pcusp_diag_VbarV} (right) meandering both horizontally and vertically.
This diagram can be reduced in a similar manner as above.   
Integrating out one row of plaquette matrices yields a factor $(1/N_c)^{2(z-1)}$, 
and the final row also yields a factor of $\Tr 1 = N_c$, so that 
\begin{equation}
\big\la\prod_r \mcP_0^{(r)} \mcP_L^{(r)\dag}\big\ra 
  = \frac{p^{q L_t L}\cdot(\tilde p_q/p^q)^{2z-1}}{N_c^q} \cdot\frac1{N_c^{2(z-1)(q-1)}}\cdot N_c
  = \frac{p^{q L_t L}\cdot(\tilde p_q/p^q)^{2z-1}}{N_c^{(2z-1)(q-1)}}\,.
\end{equation}
From the above, it follows that for \FTEE one obtains $\tilde S^{(q)}_{|Q\bar Q}=(2z-1)\log N_c$, which in agreement with the result obtained for the cusp-on-vertex 
replica geometry. As previously, this holds only when $\tilde p_q\to p^q$.

\section{Conclusions
  \label{sec:conclusions}}
We computed here the Flux Tube Entanglement Entropy (FTE$^2$) defined in \cite{Amorosso:2024leg} for arbitrary representations of static quarks in (1+1)D. 
The computation was motivated by our parallel work in \cite{Amorosso:2024leg} for (2+1)D Yang-Mills theory, where we conjectured that FTE$^2$ receives contributions in $D>2$ dimensions from vibrational and internal contributions. Since the former is absent in (1+1)D, one can extract the behavior of FTE$^2$ for the latter by a (1+1)D computation. 
We discussed the computation of FTE$^2$ on a Euclidean lattice, reviewing its formulation on both lattices with cusps on vertices~\cite{Amorosso:2024leg}, as well as on lattices with cusps in plaquettes, a construction introduced in \cite{Chen:2015kfa}. We laid out the explicit form  the density matrix and FTE$^2$ takes when quarks and antiquark sources are either in regions $V$ or $\bar{V}$.

We calculated FTE$^2$ in a variety of setups, beginning with lattices with cusps on vertices. 
We first performed a sample calculation for the color flux tube crossing both one and two boundaries. 
We introduced a diagrammatical method which allows one to  compute FTE$^2$ with cusps on vertices in (1+1)D with arbitrary number of regions and quark(antiquark) placements in $V$ or $\bar{V}$. 
This allowed us to prove for the cusp on vertex geometry,  FTE$^2$ takes the form conjectured in \cite{Amorosso:2024leg} that  ${\tilde S}_\text{internal} = F \cdot \log(\dim R_{Q(\bar Q)})$, where $F$ corresponds to the number of boundary crossings and $\dim R_{Q(\bar Q)}$ is the dimension of the irreducible color representation of the (anti)quark. 
We then computed FTE$^2$ for the lattice geometry corresponding to cusps in plaquettes in the general case, introducing a further diagrammatical scheme to calculate FTE$^2$. 
We demonstrated that FTE$^2$ in this geometry is also given by ${\tilde S}_\text{internal} = F \cdot \log(\dim R_{Q(\bar Q)})\,$, as long as the coupling of the enlarged (``big") plaquette is tuned such that its average value is the average value of a standard plaquette to the power of the number of replicas. 
Our results motivate further numerical computations of FTE$^2$ in $D>2$ (and for greater $N_c$ than the $N_c=2$ results reported in \cite{Amorosso:2024leg}) for multipartite ``slab" geometries to further isolate the contribution of internal and vibration modes to FTE$^2$.

\section*{Acknowledgements}
R.A is supported by the Simons Foundation under Award number 994318 (Simons Collaboration on Confinement and QCD
Strings). S.S. is supported by NSF supported by the National Science Foundation under award PHY-2412963. 
In addition, R.A. is supported in part by the Office of Science, Office of Nuclear Physics,
U.S. Department of Energy under Contract No. DEFG88ER41450 and by the National Science Foundation under award
PHY-2412963.
R.V is supported by the U.S. Department of Energy, Office of Science under contract DE-SC0012704. 
R.V's work on quantum information science is supported by the U.S. Department of Energy, Office of Science, National
Quantum Information Science Research Centers, Co-design Center for Quantum Advantage (C$^2$QA) under contract number
DE-SC0012704. 
R.V. was also supported at Stony Brook by the Simons Foundation as a co-PI under Award number 994318 (Simons Collaboration on Confinement and QCD Strings). 
The authors thank Stony Brook Research Computing and Cyberinfrastructure and the Institute for Advanced Computational
Science at Stony Brook University for access to the Seawulf HPC system, which was made possible by grants from the
National Science Foundation (awards 1531492 and 2215987) and matching funds from the Empire State Development’s Division
of Science, Technology and Innovation (NYSTAR) program (contract C210148).

\bibliography{bib}

\end{document}